\documentclass[aps,pra,showpacs,groupedaddress,twocolumn]{revtex4}
\usepackage{amsfonts}
\usepackage{amsmath}
\usepackage{mathrsfs}
\usepackage{amssymb}
\usepackage[dvips]{graphics,color}
\usepackage{graphicx}
\usepackage{dcolumn}
\usepackage{bm}
\begin{document}
\title{Coherent bimolecular reactions with quantum-degenerate matter waves}
\author{H. Jing$^{1}$, J. Cheng$^2$ and P. Meystre$^3$}
\affiliation{$^1$Department of Physics, Henan Normal University,
Xinxiang 453007, People's
Republic of China\\
$^2$School of Physical Science and Technology, South China
University of Technology, Guangzhou 510640,
 People's Republic of China\\
 $^3$B2 Institute and Department of Physics, The
University of Arizona, Tucson, Arizona 85721, USA}
\date{\today}
\begin{abstract}
We demonstrate theoretically that the abstraction
reaction $A+B_2 \rightarrow AB+B$ can be driven coherently and efficiently with quantum-degenerate bosonic or fermionic matter waves. We show that the initial stages of the reaction are dominated by quantum fluctuations, resulting in the appearance of macroscopic non-classical correlations in the final atomic and molecular fields. The dynamics associated with the creation of bosonic and of fermionic dimer-atom pairs are also compared. This study opens up a promising new regime of quantum degenerate matter-wave chemistry.
\end{abstract}
\maketitle

\section{Introduction}

The making and probing of ultracold molecular gases have attracted
much attention in recent years \cite{krems}, as their realization
opens up exciting applications in fields ranging from condensed
matter physics to quantum information science. The application of
magnetic Feshbach resonances (FR)
\cite{Robert,Greiner,Kerman,Danzl,FR 1} and of optical
photoassociation (PA) \cite{Paul,PA 1,PA 2,PA 3}, oftentimes in
combination, within an atomic Bose-Einstein condensate can result
in the creation of diatomic \cite{Lang} as well as of more complex
molecules, as evidenced by recent experimental observations of
transient Efimov trimer states Cs$_3$ \cite{trimer} and of the
molecular tetramer Cs$_4$ (indirectly via resonances in inelastic
processes) \cite{Efimov,tetramer}. We also mention the important
work of Ling $et~ al.$ \cite{Ling,Pu,Ling 1}, who proposed to use
a chirped coupling field to compensate the effects of nonlinear
collisions within the stimulated Raman adiabatic passage (STIRAP)
technique \cite{F,D,M,STIRAP 1,STIRAP 2,STIRAP 3,STIRAP 4} and
thus to efficiently generate large amounts of deeply bound
ultracold molecules. This is an extension of previous
developments in associative STIRAP (see, e.g. early work by
Mackie and coworkers \cite{STIRAP 4}), that is particularly relevant
in the context of the present paper. The atom-molecule
dark states involved in such a process was first realized
experimentally via coherent two-color PA \cite{Winkler} in an
early example of what is now called superchemistry \cite{Heinzen}.

Over the past ten years, ultracold matter-wave superchemistry has
focused on coherent association or dissociation reactions
\cite{PD,Bongs} between atoms and diatomic molecules. For example,
Moore and Vardi \cite{PD} studied the possibility of an almost
complete Bose-enhanced channel selectivity in the coherent
photodissociation of bosonic triatomic $ABC$ molecules, resulting
from the interplay between Bose enhancement and competition between
modes for a finite number of initial molecules.  Other recent
examples include work on the dependence of dissociation on the size
or shape of the reaction vessel (confinement effect) \cite{Vardi}
and the assembly of Fermi-degenerate dimers via cooperative
association \cite{olavi}.

The present paper extends the toolbox of superchemistry to the
coherent abstraction reaction (or bimolecular reactive scattering)
$A+B_2\rightarrow AB+B$, where $A, B$, $B_2$ and $AB$ denote either
bosonic or fermionic atoms or dimers. This reaction is an important
benchmark system in chemical physics. Its dynamics has attracted
much interest in recent studies of reactive resonance or low-energy
non-Born-Oppenheimer reactivity through a cross-beam scattering
method \cite{bodo,che}. A particularly noteworthy contribution is
the study by Shapiro and Brumer of the coherent control of
single-molecular photoassociation or bimolecular collisions through
the interference of reactive pathways \cite{disp}.

Extending these considerations to the case of ultracold matter
waves, we show that the coherent abstraction reaction can be
realized efficiently and controlled in a STIRAP photoassociation
pulse sequence \cite{Paul} such that the intermediate states are
dark states. An important characteristic of this process is that it
is triggered by quantum noise, leading to large shot-to-shot quantum
fluctuations that invalidate the use of the Gross-Pitaevskii
equation (GPE) in the initial stages. That equation can be used only
at later times when the product reactant channels become
macroscopically occupied. This is somewhat similar to a situation
familiar in a number of quantum and atom optics examples such as the
laser \cite {QO}, optical and matter-wave superradiance \cite{Uys},
and molecular matter-wave amplifier \cite{amplifier}, and is in
contrast to the familiar single-molecular \cite{Paul,Heinzen,Jing}
combination reaction.

In realizing the collective reaction $A+ B_2 \rightarrow AB+B$, the
basic idea is to first create weakly bound trimers $AB_2$ via an
entrance-channel atom-dimer FR, and then to dissociate them into a
closed-channel bound dimer and atom via photodissociation. A key
aspect of that scheme is that involving a trimer
intermediate state allows one to exploit a coherent population
trapping (CPT) state that prevents the trimer population from
becoming significant throughout the conversion process. Such an
atom-molecule state does $not$ exist in other schemes that involve
$e.g.$ an intermediate two-species atomic state. Note also that this
scheme, which is specific to quantum-degenerate matter waves, is
different from a purely collision-induced reaction \cite{3body} and
from the non-degenerate single-pair dynamics of reactive scattering
\cite{disp}. As such it represents a promising advance in the
on-going development of superchemistry \cite{tetramer,Heinzen}.

The remainder of this paper is organized as follows: Section II
discusses our model and analyzes the initial stages of the coherent bimolecular reaction. The resulting quantum fluctuations determine the initial statistical properties of the mean-field evolution that takes over once the various matter-wave fields are macroscopically populated. This is discussed in Sec. III, where we numerically compute a large number of trajectories from initial classical seeds satisfying these short-time statistics. We also review how an approximate CPT dark state required for the STIRAP pulse sequence can be achieved in the presence of mean-field shifts. Section IV briefly discusses the possible conversion of bosons to fermions. Several generalizations, including the role of population imbalance, are considered in Sec.~V. Finally Section VI is a summary and conclusion.

\section{Short-time quantum dynamics}

\begin{figure}[ht]
\includegraphics[width=8cm]{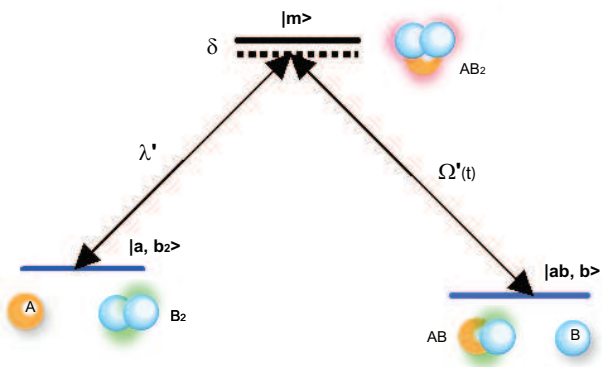}
\caption{(Color online) Schematics of the coherent
abstraction reaction $A+B_2\rightarrow AB+B$ with degenerate matter
waves. $A$ and $B$ denote the bosonic or fermionic atoms, and $B_2$
and $AB$ are molecular dimers.}
\end{figure}

Our model system is sketched in Fig.~1. The
intermediate heteronuclear trimers $AB_2$ are created via FR, and
then photodissociated into bound molecules $AB$ and atoms $B$. Denoting the atom-dimer coupling strength with detuning
$\delta$ by $\lambda'_1$, the Rabi frequency of the dissociating
laser by $\Omega'_1$ and its detuning by $\Delta$, the dynamics of
the system is described at the simplest level by the model
Hamiltonian ($\hbar=1$)
\begin{eqnarray}
\hat{\cal H}\!\!\!&=\!\!\!&-\!\!\int\!\! dr \Big \{\!\!
 \sum_{i,j}\chi_{i,j}'{\hat \psi}_i^\dag(r){\hat \psi}_j^\dag(r)
 {\hat \psi}_j(r){\hat \psi}_i(r)\!\!
 +\!\delta {\hat \psi}_m^\dag(r) {\hat \psi}_m(r) \nonumber\\
 &+&\lambda'[{\hat \psi}
 _m^\dag(r) {\hat \psi}_a(r){\hat \psi}_{b_2}(r)
 \!\!+\!{\rm H.c.}]+(\Delta+\delta){\hat \psi}_{ab}^\dag(r){\hat\psi}_{ab}(r)\nonumber\\
 &-&\Omega'[\hat{\psi}_{ab}^\dag(r){\hat \psi}_b^{\dag}(r){\hat
 \psi}_m(r)
 +{\rm H.c.}]
\Big \}.
\end{eqnarray}
We consider first a purely bosonic system. The annihilation
operators $\hat{\psi}_i(r)$, where the indices ${ i},{ j}={ a}, {
b}, { b_2}, { ab}, { m}$ stand for atoms ($A$ and $B$), dimers
($B_2$ and $AB$) and trimers ($AB_2$), satisfy the standard bosonic
commutation relations
$$[\hat{\psi}_i(r),~\hat{\psi}^\dag_j(r')]=\delta_{ij}\delta(r-r').$$
The terms proportional to $\chi'_{i,j}={2\pi a_{i,j}}/{M_{i,j}}$
describe interspecies $s$-wave collisions with scattering length
$a_{i,j}$, $M_{i,j}={M_iM_j}/{(M_i+M_j)}$ being the reduced mass
\cite{footnote1}.

In the mean-field approximation, $\hat{\psi_i}\rightarrow\sqrt{n}\psi_i$ where $n$ is the initial particle density, the Heisenberg equations of motion resulting from the Hamiltonian~(1) are easily shown to reduce to the form
\begin{eqnarray}
\dot{\psi}_{ a}&=&2i \sum_{ j} \chi_{{ a,j}} |\psi_{ j}|^2
\psi_{{ a}}+i\lambda\psi_{ b_2}^*\psi_{ m},\nonumber\\
\dot{\psi}_{ b}&=&2i \sum_{ j} \chi_{{ b,j}} |\psi_{ j}|^2 \psi_{{
b}}-i\Omega\psi_{ ab}^{*}\psi_{
m},\nonumber \\
\dot{\psi}_{ b_2}&=&2i \sum_{ j} \chi_{{ b_2,j}} |\psi_{ j}|^2
\psi_{ b_2}+i\lambda\psi_{ a}^*\psi_{ m},\nonumber\\
\dot{\psi}_{ ab}&=&2i \sum_{ j} \chi_{ ab,j} |\psi_{ j}|^2 \psi_{
ab}-i\Omega\psi_{ b}^*\psi_{ m}
+i(\Delta+\delta)\psi_{ ab},\nonumber\\
\dot{\psi}_{ m}&=&2i \sum_{ j} \chi_{{ m,j}} |\psi_{ j}|^2 \psi_{
m}+(i\delta\
-\gamma) \psi_{ m}+i\lambda\psi_{ a}\psi_{ b_2}\nonumber\\
&&-i\Omega\psi_{ b}\psi_{ ab}.
\end{eqnarray}
Here $\chi_{ i,j}=n\chi_{ i,j}',~\lambda=\lambda'\sqrt{n}$,
$\Omega=\Omega'\sqrt{n}$, and we have introduced the
phenomenological decay rate $\gamma$ to account for the loss of
intermediate trimers, based on the assumption that this decay
dominates all other loss mechanisms such as rogue photodissociation
to noncondensate modes \cite{FR 1,rogue1}. As already mentioned, our
goal is to minimize that decay by using a STIRAP pulse sequence,
ideally permitting the full transfer of the entrance-channel state
to the closed-channel state while keeping the intermediate state
unpopulated at all times.

The initial condition ${\psi}_b(0) = \psi_{ab}(0)=0$ is readily
seen to result in $${\psi}_b(t) = \psi_{ab}(t)=0,$$ for all times.
This indicates that the mean-field GP equations break down
completely in studying the onset of this type of abstraction
reaction. A similar situation has been previously
encountered in a broad range of systems in quantum optics
\cite{QO}, but also in coupled degenerate atomic and molecular
systems such as the example in the matter-wave superradiance of
Bose-condensed atoms \cite{QO,Uys,amplifier}. As in those
situations, our strategy here is to decompose the problem into an
initial quantum-noise-dominated stage followed by a classical
stage that arises once the product components have acquired a
macroscopic population. The initial quantum evolution is treated
in a linearized approach whose main purpose is to establish the
statistical properties of the initial fields required for the
classical stage \cite{amplifier}.

A simple physical picture of the initial stages of the coherent
abstraction reaction can be obtained when considering the limiting
case where $\delta$ is the largest parameter in the system,
$i\dot{\hat{\psi}}_m/\delta \thickapprox 0$. In the collisionless
limit this gives
$$\hat{\psi}_m\thickapprox
-(\lambda'/\delta)\hat{\psi}_a\hat{\psi}_{b_2}+(\Omega'/\delta)\hat{\psi}_b\hat{\psi}_{ab},$$
which amounts to adiabatically eliminating the intermediate
trimer state. In this case the system is described by the effective Hamiltonian
\begin{equation}
\hat{\cal H}_{\rm eff}=-(G \hat{c}_{ab}^\dag \hat{c}_b^\dag
\hat{c}_a \hat{c}_{b_2}+{h.c.})+\hat{c}_0,
\end{equation}
where
\begin{equation}
\hat{c}_0=\omega_1\hat{c}_a^\dag \hat{c}_{ a} \hat{c}_{
b_2}^\dag \hat{c}_{ b_2} +\omega_2 \hat{c}_{ab}^\dag \hat{c}_{ ab}
\hat{c}_{ b}^\dag \hat{c}_{ b},
\end{equation}
$$
\hat{\psi}_{ i}(r,t)=\phi_{ i}(r)\hat{c}_{ i}(t),
$$
and the various constants are
    \begin{eqnarray}
    G&=&(\lambda'\Omega'/\delta)\int dr \phi_{ ab}^*(r) \phi_{
    b}^* (r)\phi_{ a}(r) \phi_{ b_2}(r),\nonumber \\
    \omega_1&=&(\lambda'^2/\delta)\int dr \phi_{ a}^* (r)\phi_{
    a} (r)\phi_{ b_2}^* (r)\phi_{ b_2}(r),\nonumber \\
    \omega_2&=&(\Omega'^2/\delta)\int dr \phi_{ ab}^* (r)\phi_{
    ab}(r) \phi_{ b}^* (r)\phi_{ b}(r),\nonumber
    \end{eqnarray}
The Hamiltonian (3), which is exactly solvable, has been considered previously in the study of a spin exchange scattering process that produces entangled bosonic pairs in a two-species, two-pseudospins Bose condensate \cite{Wu}.

For short enough interaction times, Eq.~(3) can be further
simplified by taking into account the fact that the populations of
the reaction products remain small compared to the total particle
numbers $N_0$. In this regime, we can treat the fields ${\hat
\psi}_{ a}$ and ${\hat \psi}_{ b_2}$ classically, $\hat{c}_{ a,
b_2}\rightarrow \sqrt{N_{ a,b_2}}$, and then neglect the term in
Eq.~(3) describing only the interactions between the modes
$\hat{c}_{ a}$ and $\hat {b}_2$. As discussed e.g. in Ref.
\cite{Uys}, this results in the linearization of the Heisenberg
equations of motion for the remaining quantized matter-wave fields
${\hat c}_{ab}$ and ${\hat c}_b$, with a noise source ${\hat
f}_j^\dagger(t)$,
\begin{equation}\label{4}
\dot{{\hat c}}_{ab,b}(t)={\hat f}^\dag_{b,ab}(t)=i{\cal G}{\hat
c}_{b,ab}^\dag(t).
\end{equation}
Such a result is familiar from several quantum and atom optics
problems, including the optical parametric oscillator \cite{QO} and
molecular dissociation (pair production) in ultracold systems
\cite{twin}. For simplicity we assume in the following that the
correlations of the quantum noise operators appearing in
Eq.~(\ref{4}) are markovian,
\begin{equation}\label{5}
\langle {\hat f}_{ i}^\dagger(t){\hat f}_{ j}(t') \rangle=0,~~
\langle {\hat f}_{ i}(t){\hat f}_{ j}^\dagger(t')\rangle ={\cal
G}^2\delta_{ij}\delta(t-t'),\nonumber
\end{equation}
where ${\cal G}=G\sqrt{N_{ a}N_{ b_2}}$ and $i,j=ab$ or $b$ here and
in the following. It is these noise operators that trigger the
non-mean-field "spontaneous" evolution of the system from initial
vacuum fluctuations.

The populations of the modes $j=b, ab$ and their correlations are then
\begin{eqnarray}
    N_j&\equiv& \langle \hat{c}_j^\dagger\hat{c}_j \rangle
    =\sinh^2({\cal G}t) \approx N_aN_{b_2}G^2t^2; \nonumber\\
    {\cal C}_{ab}={\cal C}_b &\equiv& \frac{\langle \Delta
    \hat{N}_{ab}\Delta \hat{N}_b \rangle}{\sqrt{N_{ab}N_b}}=1+\sinh^2({\cal G}t)>1,
\end{eqnarray}
where $\Delta \hat{N}_j\equiv\hat{N}_j-\langle \hat{N}_j \rangle$.
Eq. (5) can also be derived by solving Eq. (3) to second order in
time $t$ with the depletions
$N_{\mathrm{a},\mathrm{b_2}}$$=N_0-N_{\mathrm{ab},\mathrm{b}}$. It is also straightforward to find that
$$
\langle
\hat{N}_{ab}^2\rangle=\langle\hat{N}_b^2\rangle=\langle\hat
{N}_{ab}\hat{N}_b\rangle=\sinh^2({\cal G}t)\cosh^2({2\cal G}t)
$$
and that
\begin{eqnarray}
g^{(2)}_b&=&g^{(2)}_{ab}=\frac{\langle
\hat{c}^\dag_{ab}\hat{c}^\dag_{ab}\hat{c}_{ab}\hat{c}_{ab}\rangle}{\langle
\hat {N}_{ab}\rangle^2}=2.
\end{eqnarray}
It follows that the Mandel $Q$ parameters \cite{QO} is given by
\begin{equation}
Q_{ab,b}=\frac{\langle \hat{N}^2_{ab,b} \rangle- \hat{N}_{ab,b}^2}{
\hat{N}_{ab,b}}= ~\cosh^2({\cal G}t)>1,
\end{equation}
and exhibits super-Poisson statistics \cite{QO}. It is interesting
to observe that although the second factorial moments of the single
modes $ab$ and $b$ are typical of chaotic fields, quantum
entanglement within these two modes does exist, i.e.,
\begin{eqnarray}
&&g^{(2)}_{ab,b}=\frac{\langle
\hat{N}_{ab}\hat{N}_b\rangle}{N_{ab}N_b}=1+\frac{\cosh^2({\cal
G}t)}{\sinh^2({\cal G}t)},\\
&&\left [g_{ab,b}^{(2)}\right
]^2-g_{ab}^{(2)}g_b^{(2)}=\sinh^{-4}({\cal G}t)+4\sinh^{-2}({\cal
G}t)>0 \nonumber,
\end{eqnarray}
violating the classical Cauchy-Schwarz inequality (CSI) \cite{QO,
footnote2}.

Similar equations of motion can be derived in case atoms $A$ are
bosonic and atoms $B$ fermionic. The main difference in that case
is in the commutation relations of the  noise operators
$(-\hat{f}^\dag_{b},~\hat{f}^\dag_{ab})$. The Heisenberg equations
of motion can be solved via a Bogoliubov transformation. One finds
that the vacuum-noise-triggered populations of principal modes are
then \cite{twin}
$$
N_{ab,b}=\sin^2({\cal G}t)<1,
$$
a direct consequence of the Fermi statistics \cite{twin}. Being
similar to the  bosonic case, we also find
$\langle\hat{N}_{ab}^2\rangle=\langle
\hat{N}_{b}^2\rangle=\sin^4({\cal G}t)+\cos^2({\cal
G}t)\sin^2({\cal G}t)=\sin^2({\cal G}t)$, $\langle
\hat{N}_{ab}\hat{N}_b\rangle=-\sin^2({\cal G}t)\cos(2{\cal G}t))$,
and
\begin{eqnarray}
&&g^{(2)}_{ab}= g^{(2)}_{b}=\frac{\langle
\hat{c}^\dag_{ab}\hat{c}^\dag_{ab}\hat{c}_{ab}\hat{c}_{ab}\rangle}{\langle
\hat {N}_{ab}\rangle^2}=0, \nonumber\\
&&g^2_{ab,b}=\frac{\langle\hat{N}_{ab}\hat{N}_b\rangle}{
N_{ab}N_b}=1-\frac{\cos^2({\cal G}t)}{\sin^2({\cal
G}t)},\nonumber\\
&&\left [g^{(2)}_{ab,b}\right ]^2-g^{(2)}_{ab}g^{(2)}_b=
[1-\frac{\cos^2({\cal G}t)}{\sin^2({\cal G}t)} ]^2>0.
\end{eqnarray}
The fermionic dimer-atom pairs correlations are of course also
different from the bosonic case, specifically we have now
$C_{ab,b}=1-N_{ab,b}<1$, a signature of antibunching.
Additionally, the Mandel $Q$ parameter for the principal mode is
\begin{equation}
Q_{ab,b}=\frac{\langle \hat{N}^2_{ab,b} \rangle-
\hat{N}_{ab,b}^2}{\hat{N}_{ab,b}} = ~\cos^2({\cal G}t)<1,
\end{equation}
characteristic of subpoissonian statistics.

\section{Long-time classical evolution}

The long-time statistical properties of the $AB$ and $B$
populations, which are significantly influenced by the initial
vacuum fluctuations, can be calculated by a positive-$P$
representation technique \cite{Heinzen} and other methods
\cite{Wu}. Rather than adopting such a full quantum treatment, we
proceed in the following by solving the mean-field description of
Eqs.~(2) with stochastic classical seeds whose statistics are
consistent with the results of the linearized, short-time quantum
analysis \cite{amplifier}. To be specific, using Eqs. (2) we
compute $300$ trajectories with randomly chosen initial classical
seeds satisfying the short-time behavior of Eq.~(5).

Figure~2 shows the standard derivations $\Delta N_i(t)$ around the
average values of the particle populations,
$$ \Delta N_i(t)=\left \{\frac{1}{300} \sum_{n=1}^{300}
\left [(N_{i,n}(t)-\bar{N}_{i}(t)\right ]^2 \right \}^{1/2},
$$
with $\bar{N}_i(t)=(1/300)\sum_nN_{i,n}(t),$ for $\delta=3$
and $\delta=-3$. The inset shows the fluctuating range
$\pm \Delta N_i$ about the mean populations, $\bar{N}_i\pm \Delta N_{i}(t)$
for $\delta=3$ and for bosonic atoms. The small seeds resulting
from the initial quantum fluctuations are
significantly amplified, increasing more rapidly than their
deviations, before reaching a stationary value. For $\delta=-3$,
however, no stable reaction is observed.

\begin{figure}[ht]
\includegraphics[width=8cm]{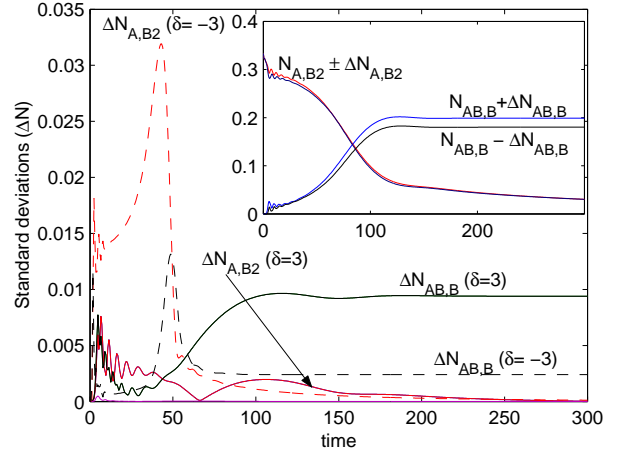} \caption{(Color online)
Standard derivation of the dimer and atomic populations
from their average values for $\delta=3$ and $\delta =-3$. Time is in
units of $\lambda^{-1}$, and $\gamma =1$. The other parameters are
given in the text. The trimer population remains essentially zero at all times due to the CPT condition. Inset: fluctuating range of the populations
$\bar{N}_i(t)\pm \Delta N_i(t)$ for $\delta=3$.}
\end{figure}

An important feature of the coherent abstraction reaction is that it
can be controlled and optimized by exploiting the existence of a CPT
dark state \cite{Winkler,Jing}. This technique is well known in the
case of linear systems, where it permits the transfer of population
from an initial to a final state via an intermediate state that
remains unpopulated at all times. This is the basis for stimulated
Raman adiabatic passage (STIRAP), which achieves this goal via a
so-called counter-intuitive sequence of pulses \cite{Winkler,STIRAP
5,STIRAP 6,STIRAP 7}.

CPT and STIRAP rely explicitly on the validity of the adiabatic
theorem, which applies only to linear systems. While there have
been many previous developments in associative STIRAP (see e.g.
Ref. [26]), it is not immediately obvious that STIRAP still works in the usual way with a well-defined nonlinear adiabatic condition for the nonlinear
system at hand, even in the collisionless limit.
This question was recently investigated by Pu and coworkers in
Ref. [16] for a model of coherent atom-dimer conversion, and an approximate adiabatic condition was obtained by linearizing the
nonlinear system around the intended adiabatic evolution. In that
context, the analytical form of the nonlinear adiabatic condition
derived in the collisionless limit turns out to be useful in
determining the required laser parameters \cite{Pu,nonlinear
1,nonlinear 2,Ling 1}.

We show in the following that an approximate atom-molecule CPT
state can also be achieved in the present situation. Specifically,
Eqs.~(2) admit a steady-state CPT solution with a trimer state
that remains unpopulated at all times under the generalized
``two-photon" resonance condition
\begin{align}
\Delta=-&\delta+(2\chi_{{ aa}}+6\chi_{{ab}}+5\chi_{{
bb_{2}}}+2\chi_{{ bb}})
N_{{b_{2},s}}\nonumber\\
+&(2\chi_{{ ab}}+\chi_{{ bb_{2}}}) N_{{ ab,s}},
\end{align}
To this end, we apply the steady-state ansatz
\begin{align}
\psi_{a,s}&=|\psi_{a,s}|e^{i\theta_{a}}e^{i\mu_{a}t},~~
\psi_{b,s}=|\psi_{b,s}|e^{i\theta_{b}}e^{i\mu_{b}t},\nonumber\\
\psi_{b_{2},s}&=|\psi_{b_{2},s}|e^{2i\theta_{b}}e^{2i\mu_{b}t},\nonumber\\
\psi_{ab,s}&=|\psi_{ab,s}|e^{i(\theta_{a}+\theta_{b})}e^{i(\mu_{a}+\mu_{b})t},\nonumber\\
\psi_{m,s}&=|\psi_{m,s}|e^{i(\theta_{a}+2\theta_{b})}e^{i(\mu_{a}+2\mu_{b})t},
\end{align}
where $\mu_a$ and $\mu_b$ are the atomic chemical potentials.
Inserting these trial functions into Eqs. (2) and taking
$|\psi_{m,s}|=0$, one finds the steady-state relation
$$\lambda\psi_{a,s}\psi_{b_{2},s}=\Omega\psi_{b,s}\psi_{ab,s},$$
which, together with the condition of conserved particle numbers
$N_{a,s}+2(N_{b_{2},s}+N_{ab,s})+N_{b,s}=1$, gives
\begin{equation}
\bigl[(\frac{\Omega}{\lambda})-1\bigl] N_{ab,s}^{2}+ \frac{2{\cal
R}+1}{2{\cal R}+2}N_{ab,s}-\frac{{\cal R}}{2({\cal R}+1)}=0,
\end{equation}
where
\begin{align}
{\cal R}\equiv \frac{{N_{{a}}(0)}}{{2N_{{b_{2}}}(0)}}=
\frac{N_{a,s}+N_{ab,s}}{N_{b,s}+N_{ab,s}+2N_{b_{2},s}}.
\end{align}
The CPT solution is therefore
\begin{eqnarray}
N_{{ab,b}}^s=\frac{2{\cal R}}{(1+{\cal R})\bigl[1+2{\cal
R}+\sqrt{(1-2{\cal R})^{2}+8{\cal
R}\Omega^{2}/\lambda^{2}}\bigl]}.\nonumber
\end{eqnarray}
Table 1 displays the CPT particle numbers for several values of
${\cal R}$. Note that from the conservation of particle numbers we
have $N_{b,s}=N_{ab,s}$ and
$$N_{{b_{2},s}}=\frac{1}{{2(1+{\cal
R})}}-N_{{ab,s}}.$$ In addition, we note that an initial
populations imbalance can also significantly affect the dynamics
of atom-molecule conversion for our present bimolecular reactions.
The condition ${\partial N_{{ ab,s}}}/{\partial {\cal R}}=0$
yields a maximum dimer number
$$
N_{{ ab,s}}|_{\max}= {1}/{3},
$$
corresponding to a complete abstraction reaction for ${\cal
R}={1}/{2}$ or the so-called "balanced case". We
consider this case first and then turn to study the effect
of initial populations imbalance (see Section V).
\begin{table}
\centering
\begin{tabular}{l|lll}
  \hline\hline
  ${\cal R}$ & $N_{g_2,s;b,s}$ & $N_{g_1,s}$ & $N_{a,s}$ \\
  \hline
 1& $\frac{1}{3+\Omega_{eff}}$ & $\frac{2\Omega^2/\lambda^2}{(3+\Omega_{eff})(1+\Omega_{eff})}$
 & $\frac{1+\Omega_{eff}+4\Omega^2/\lambda^2}{(3+\Omega_{eff})(1+\Omega_{eff})}$ \\
 $\frac{1}{2}$ & $\frac{1}{3(1+\Omega/\lambda)}$
    & $\frac{\Omega/\lambda}{3(1+\Omega/\lambda)}$ & $\frac{\Omega/\lambda}{3(1+\Omega/\lambda)}$ \\
 $\frac{1}{4}$ & $\frac{4}{5(3+\Omega_{eff})}$ &
 $\frac{2(1+\Omega_{eff})}{5(3+\Omega_{eff})}$
 & $\frac{8\Omega^2/\lambda^2}{5(3+\Omega_{eff})(1+\Omega_{eff})}$ \\
 \hline
\end{tabular}
  \caption{Steady-state CPT particle numbers, with  $\Omega_{eff}=\sqrt{1+8\Omega^2/\lambda^2}$.}\label{1}
\end{table}

We have numerically solved Eqs.~(2) and the typical results are
showed in Fig. 3. In this specific example atom A is $^{87}$Rb,
atom B is $^{41}$K, $\lambda= 4.718\times10^4 $s$^{-1}$ and
$$
\Omega(t)=\Omega_{0}{\rm sech}(t/\tau),
$$
with $\Omega_{0}/\lambda=20$ and $\lambda\tau$=$20$ \cite{lens}.
The collision parameters, in units of $\lambda/n$, are
$\chi_{aa}=0.5303$, $\chi_{bb} = 0.3214$, $\chi_{ab}=0.8731$, all
others being equal to 0.0938 \cite{lens}. As mentioned earlier,
we have neglected rogue photodissociation to noncondensate
modes, which is proved through our direct calculations to be a
safe approximation for the present parameters of our model
\cite{scattere length,rogue2,rogue3}. On the other hand, we note
that the scattering lengths of the various particles collisions,
especially those involving molecular dimers or even some trimers,
depend on the details of the interatomic potential, are yet not
known. However, in our numerical calculations it is
straightforward to use a large set of plausible collision
parameters for the Rb-K, Rb-Na or other alkali atomic samples. We
actually have done this and found essentially the similar result
as Fig.~3: stable bimolecular conversion is always possible for
appropriate values of the external field detuning $\delta$, which
is independent of the precise collision values. This result finds its origin
in the underlying mechanism of ``generalized STIRAP,' first proposed by Ling ${\it et~ al.}$ \cite{Ling} in the context of atom-dimer conversion, and
according to which collisions need not limit the conversion
rate as long as one chooses an adiabatic passage route
that compensates for the collisional mean-field phase shifts (see Eq. (12)).

Figure~3 shows the creation of $AB$ and $B$ for $\delta=\pm3$ and
$\delta=\pm1$. The results are essentially the same as those of
Ref. \cite{Jing} for $\delta=\pm3$. For this bosonic system,
stable bimolecular conversion is always possible for negative
detunings, but the system can be unstable for positive detunings
(see Fig. 3). The increasing departure of the product populations
from the ideal CPT line is due to the fact that only an
approximate adiabatic condition exists for the CPT state.

We also analytically derived the adiabaticity parameter introduced
by Pu ${\it et~ al.}$ in the collisionless limit \cite{Pu, later},
$$
\gamma_{\mathrm{nl}}(t)\approx
\frac{|\dot{\eta}|}{1+\eta}\frac{1}{4\lambda}\ll 1,
$$
where $\eta=\lambda/\Omega$. This expression differs from that for the
corresponding linear system \cite{Pu} in that in the latter case
$\eta$ is replaced by $\eta^2$ in the denominator. Hence
adiabaticity becomes increasingly difficult to maintain in the
final stages of the STIRAP process, similarly to the case of
atom-dimer conversion \cite{Pu}.

\begin{figure}[ht]
\includegraphics[width=8.5cm]{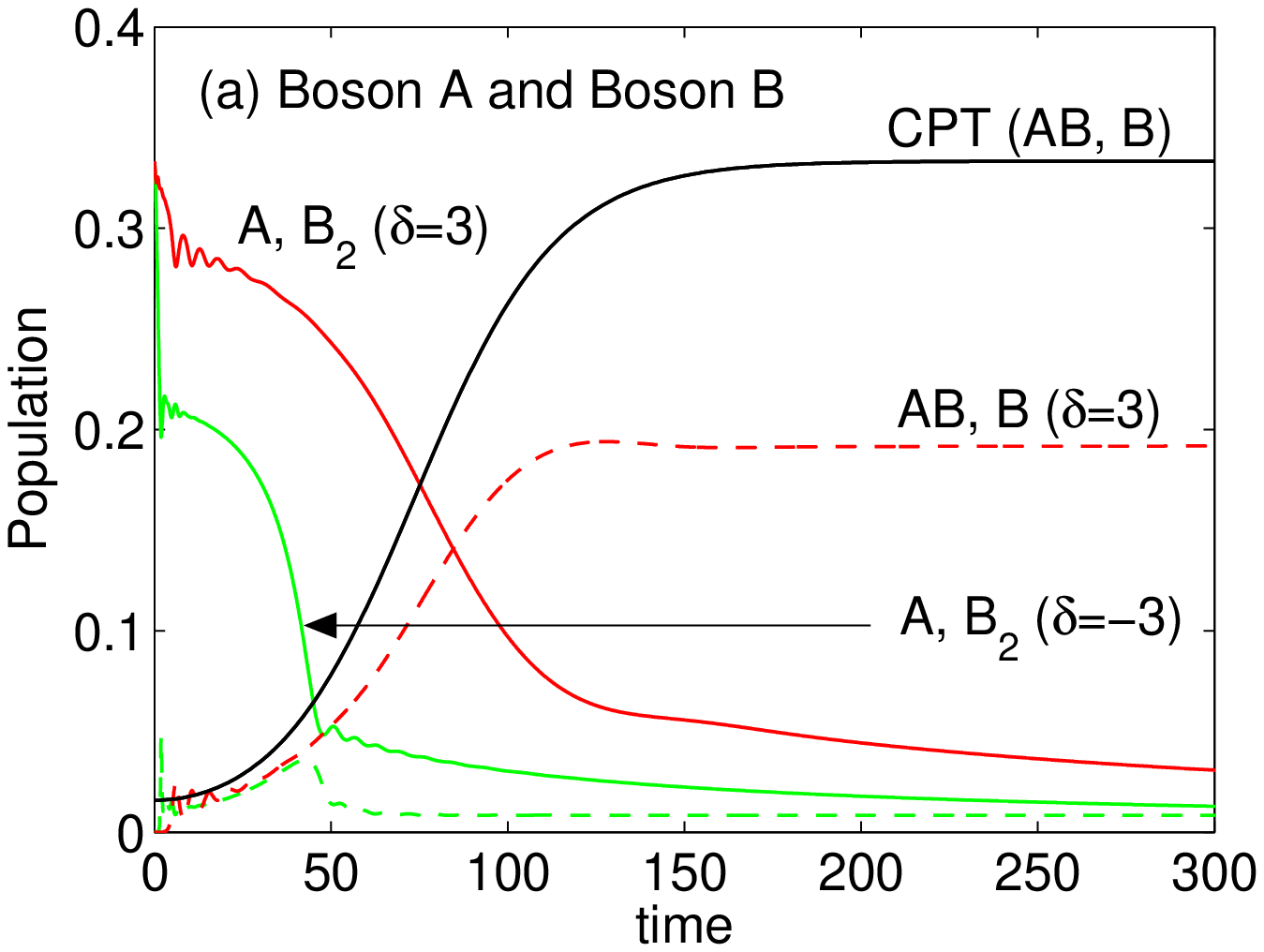}
\includegraphics[width=8.5cm]{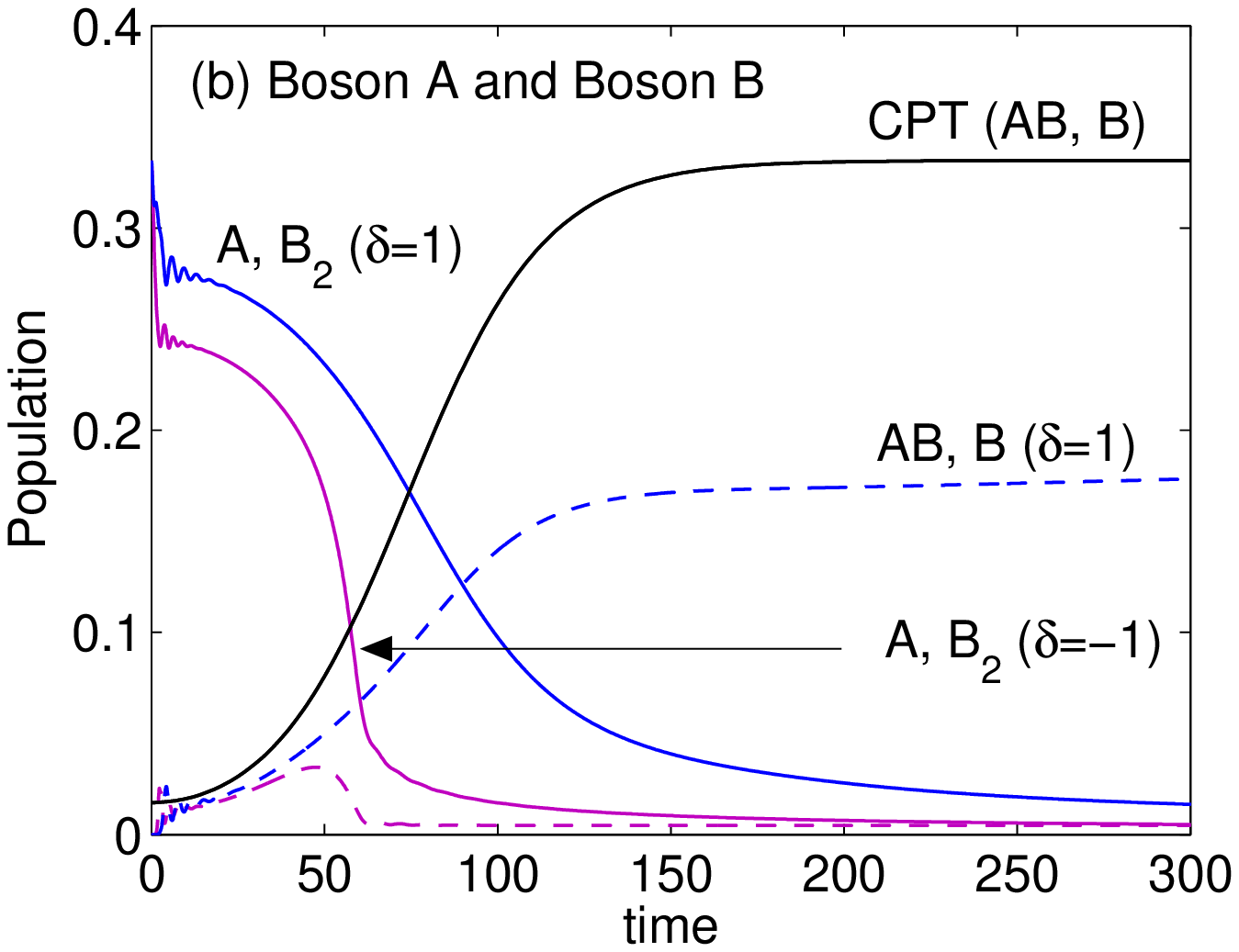}
\caption{(Color online) Populations of the bosonic dimers and atoms
for (a) $\delta=\pm3$ or (b) $\delta =\pm1$. Time is in units of
$\lambda^{-1}$, and $\gamma =1$. The line labelled ``CPT" shows the
ideal population of products (dimers AB and atoms B).}
\end{figure}

We remark that this scheme relies crucially on the capability to avoid rapid
collisional quenching or the formation of an unstable atom-dimer
sample. When energetically allowed, collision-induced reactions
always occur at some rate, and we need to guarantee that the time
scale over which quantum fluctuations dominate the dynamics is short
enough, so that the dynamics of the system is not
collision-dominated.

In order to estimate the upper limit on collisions, we can use the
condition $|{\cal G}t|<1$ or $$|G\sqrt{N_{a}N_{b_2}}t|<1,$$ which
determines the validity of the short-time approximation for the
early quantum stage, to estimate the time over which the
fluctuations take place. This upper time limit is of the order of
$10^{-5}$s for $|\delta=3|$ and $\Omega_0=20\lambda$. According to
Cvitas $et~al$. \cite{quench}, typical low-temperature inelastic
collision cross-sections are of the order of $10^{-17}m^3/s$,
corresponding to reaction times of the order of $10^{-3}s$ for a
typical condensate density of $10^{14}$cm$^{-3}$. From that
estimate, it appears that the fluctuations do indeed dominate for
short enough times. We also note that the collisional reaction
time of 10$^{-3}$s corresponds to an almost complete
noise-amplified conversion in Fig.~3. In that case the
fluctuation-induced dynamics completely dominate the short-time
behavior of the system. This feature, which is characteristic of a
wide variety of collective abstraction reactions, may provide a
useful means to produce reaction products that are difficult to
obtain or have only poor yield when resulting from a purely
collisional method.

\section{Bosons to Fermions Conversion}

When considering a mixture of bosonic and fermionic atoms, the
abstraction reaction results in the conversion of bosonic to
fermionic molecules, $$\mathbf{{b}}+\mathcal{B}\rightarrow
\mathcal{F}+\mathbf{f},$$ where $\mathbf{b}$ ($\mathcal{B}$) or
$\mathbf{f}$ ($\mathcal{F}$) denotes bosonic or fermionic atoms
(dimers). Ignoring $s$-wave collisions between fermionic atoms of
the same species and retaining only their dominant kinetic energy,
and assuming further that collisions are the dominating term for
the bosons \cite{LuLi}, the Hartree energy density of the system
is
\begin{eqnarray}
E&=& \sum_{i\neq j} \chi'_{i,j}|\psi_i|^2|\psi_j|^2
+\delta|\psi_{d}|^2
+(\Delta+\delta)|\psi_{ab}|^2 \nonumber\\
&+&\lambda'\bigl[\psi^*_{d}\psi_a\psi_{b_2}+h.c.\bigl]-\Omega'\bigl[\psi^*_{ab}\psi^*_{b}\psi_d+h.c\bigl] \nonumber\\
&+&\sum_{i=(a,b_2,d)}\frac{1}{2}\chi'_i|\psi_i|^4+\sum_{f=(ab,b)}\frac{3}{5}A'_f|\psi_f|^{10/3}.
\end{eqnarray}
Here the indices $i, j$ have the same meaning as in Eq. (1) and
$A'_f={(6\pi^2)^{2/3}}/{2{M}_f}$, where $M_f$ ($f=ab,b$) is the
mass of the fermionic components.

Due to the fermionic components of the reaction partners, the ordinary mean-field approach \cite{Parkin,Salerno} adopted in studying systems with large numbers of condensed bosonic particles is inadequate here. We follow instead the approach of Ref. \cite{LuLi}, starting from the mean-field
Lagrangian density of the system
\begin{equation}
\mathcal{L}=\displaystyle\frac{i}{2}\sum_i\Bigl(\psi^*_i\frac{\partial\psi_i}{
\partial t}-\psi_i\frac{\partial\psi_i^*}{\partial t}\Bigl)-E
\end{equation}
and exploiting the Euler-Lagrange equations
\begin{equation}
\frac{\partial\mathcal{L}}{\partial\psi_i^*}-\partial_\mu\Bigl(\frac{\partial\mathcal{L}}
{\partial(\partial_\mu\psi_i^*)}\Bigl)=0.
\end{equation}
to derive the mean-field dynamical equations
\begin{align}
\dot{\psi}_{{ a}}=&2i \sum_{ j} \chi_{{ a,j}} |\psi_{ j}|^2
\psi_{{ a}}+i\lambda\psi_{{ b_{2}}}^{*}\psi_{{ d}};\nonumber\\
\dot{\psi}_{{ b}}=&2i \sum_{ j\neq b} \chi_{{ b,j}} |\psi_{ j}|^2
\psi_{{ b}}+iA_b|\psi_{b}|^{4/3}\psi_{b}-i\Omega\psi_{{
{a b}}}^{*}\psi_{{ d}},\nonumber\\
\dot{\psi}_{{ b_2}}=&2i \sum_{ j} \chi_{{ b_2,j}} |\psi_{ j}|^2
\psi_{{ b_2}}+i\lambda\psi_{{
a}}^{*}\psi_{{ d}} \nonumber\\
\dot{\psi}_{{ ab}}=&2i \sum_{ j\neq ab} \chi_{{ ab,j}} |\psi_{
j}|^2 \psi_{{
a b}}+iA_{ab}|\psi_{ab}|^{4/3}\psi_{ab}\nonumber\\
& -i\Omega\psi_{{
b}}^{*}\psi_{{ d}}+i(\Delta+\delta)\psi_{{ ab}},\nonumber\\
\dot{\psi}_{{ d}}=&2i \sum_{ j} \chi_{{ d,j}} |\psi_{ j}|^2
\psi_{{ m}}+(i\delta\ -\gamma )\psi_{{ m}} +i\lambda\psi_{{
a}}\psi_{{ b_{2}}}.
\end{align}
These equations are similar to Eqs.~(2) with the
substitution
\begin{equation}\label{9}
\chi_{j,j}|\psi_{j}|^{2} \rightarrow A_j|\psi_{j}|^{4/3},
\end{equation}
a consequence of the fact that we consider only the dominating
kinetic energy term ignore $s$-wave collisions between identical
fermionic particles \cite{LuLi}.

In the CPT regime, the steady-state number of fermionic species
$AB$ and $B$ is therefore in the same form as in the purely
bosonic case, see Table I. However, the generalized ``two-photon"
resonance condition is now written as
\begin{align}
\Delta&=-\delta+2(\chi_{{\rm ab}}+\chi_{{\rm aa}}+\chi_{{\rm
ab_{2}}}) N_{{\rm b_{2},s}}+4\chi_{{\rm ab}} N_{{\rm
a b,s}}\nonumber\\
&+(A_\mathrm{b}-A_{\mathrm{ab}})N_{{\rm ab,s}}^{2/3}.
\end{align}

\begin{figure}[ht]
\includegraphics[width=8.5cm]{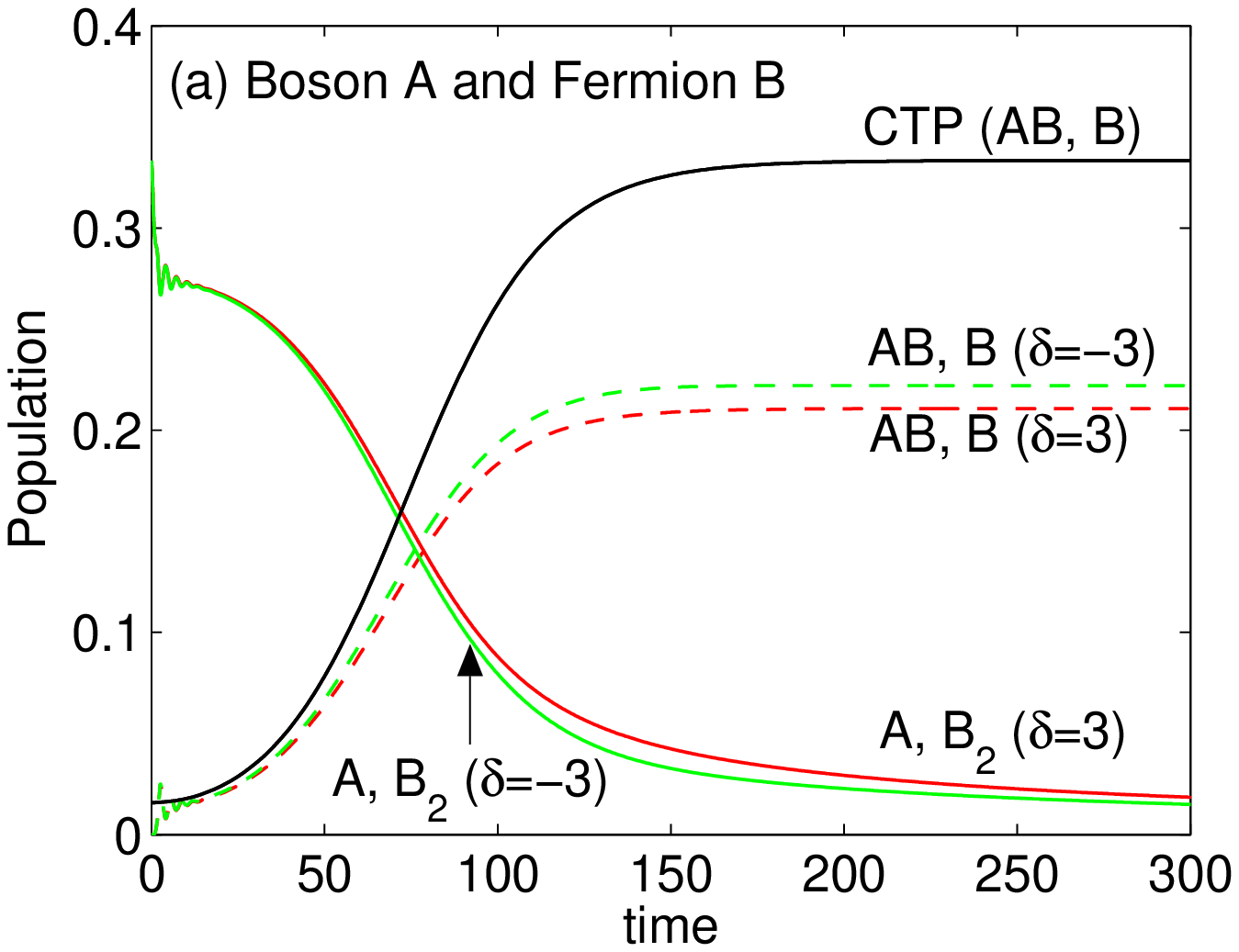}
\includegraphics[width=8.5cm]{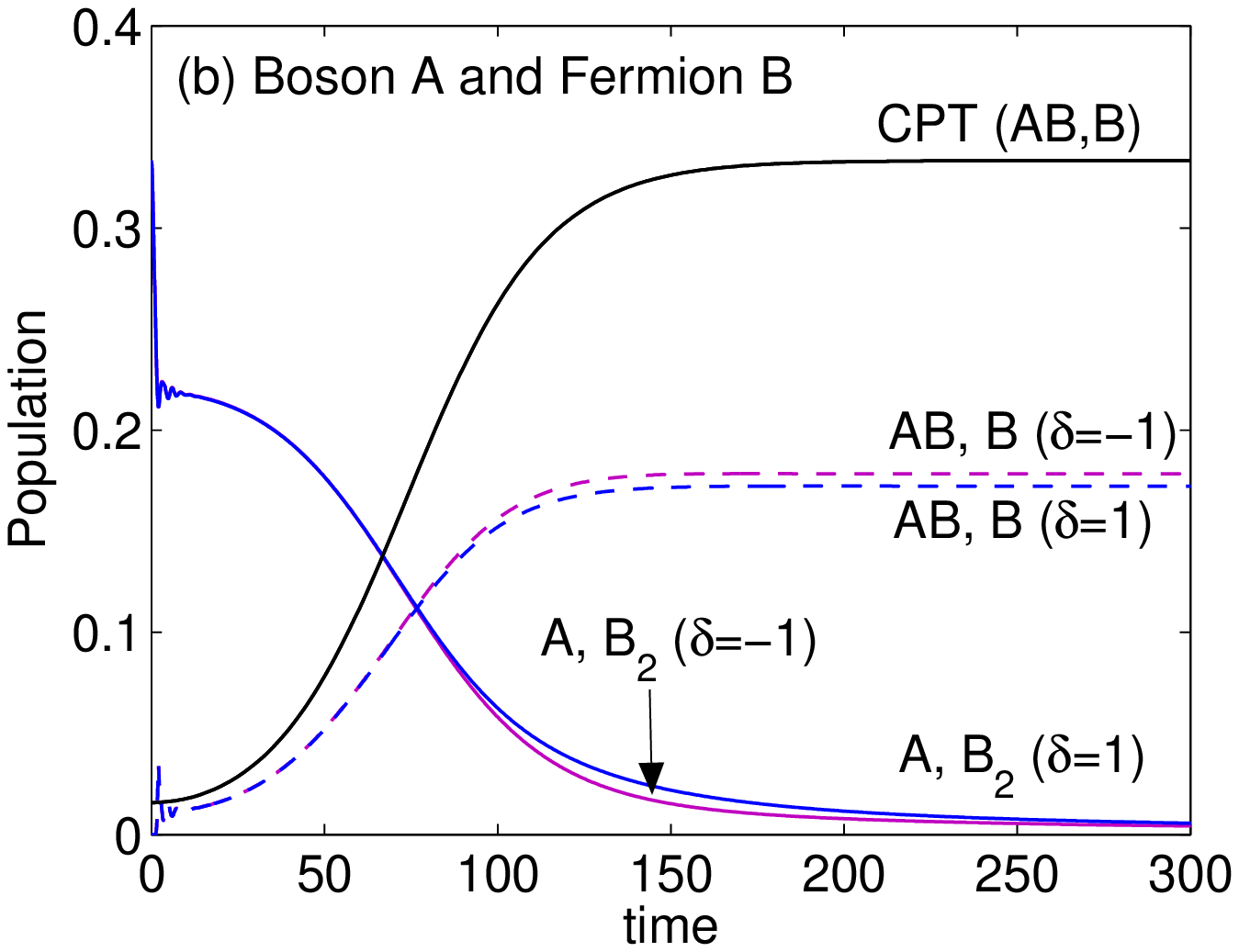}
\caption{(Color online) The generation of fermionic dimers $AB$ and
atoms $B$ for (a) $\delta=\pm3$ or (b) $\delta =\pm1$ initially with
no particle populations imbalance. The CPT values of the dimer $AB$
and atomic $B$ are also plotted.}
\end{figure}

Our numerical simulations of this case are summarized in Fig.~4
for the collision parameters $\chi_{{aa}}=0.5303$, $\chi_{{bb}}
=0$, $\chi_{{ab}}=\chi_{{b,ab}}=\chi_{{a,ab}}=-0.09$,
$\chi_{{a,ab}}=-0.2637$, $A_b=0.008$, $A_{ab}=0.004$, and all
other collision parameters equal to zero. These results show that
the stable formation of dimers $AB$ is possible for both positive
and negative detunings ((a)$\delta=\pm3$ or (b)$\delta =\pm1$),
initially with no particle populations imbalance.

We note that in contrast to the purely bosonic case, the creation
of fermion-fermion pairs is due to a statistics-independent
cooperating many-body effect that has been previously recognized
in the case of, e.g., matter-wave four-wave mixing \cite{meystre}.
We also note that in recent work Li {\it et al.} \cite{08} used a similar atom-molecule dark-state technique to realize
a so-called laser-catalyzed bimolecular reaction (or the
conversion of fermionic to bosonic molecules):
$$^6\mathrm{Li}+^6\mathrm{Li}^7\mathrm{Li} \rightarrow ^6\mathrm{Li}_2 +
^7\mathrm{Li},$$ with an ultrahigh conversion rate of 99.97$\%$
\cite{08}.

\section{Role of population imbalance}

Population imbalance often plays an important role in the physics of
ultracold matter waves. For example, using a two-spin-state mixture
of ultracold fermionic atoms, population imbalance can induce a
superfluid to normal phase transition \cite{zwier}. In the following
we demonstrate that an initial population imbalance can also
significantly affect the dynamics of this coherent collective
abstraction reaction. To this end we plot the conversion rate as a
function of ${\cal R}$, see Eq.~(14). Section III showed that in the
collisionless limit $N_{ab,s}$ reaches its maximum for ${\cal
R}=0.5$. This value is modified slightly when taking into account of
the particle collisions and the decay rate $\gamma$ in our numerical
simulations.

In Fig. 5, we see that the conversion rate
$|\psi_{ab}(t=\infty)|^2$ now has the maximum at the value
somewhat larger than $R=0.5$ for the both bosonic and fermionic
cases. In addition, we observe that the initial population
imbalance has different effects for bosons and for fermions: for
bosonic atoms $A$ and $B$, the final conversion rates can be
changed quite sharply with ${\cal R}$ and rapidly approach zero
for ${\cal R}<0.4$ and ${\cal R}>1.4$; in contrast, for the case
of fermionic atoms $B$, the conversion rate changes its shape more
slowly with ${\cal R}$ (even for ${\cal R}>1.4$, the occupation of
the product species is still in excess of 7\%).

\begin{figure}[ht]
\includegraphics[width=8.5cm]{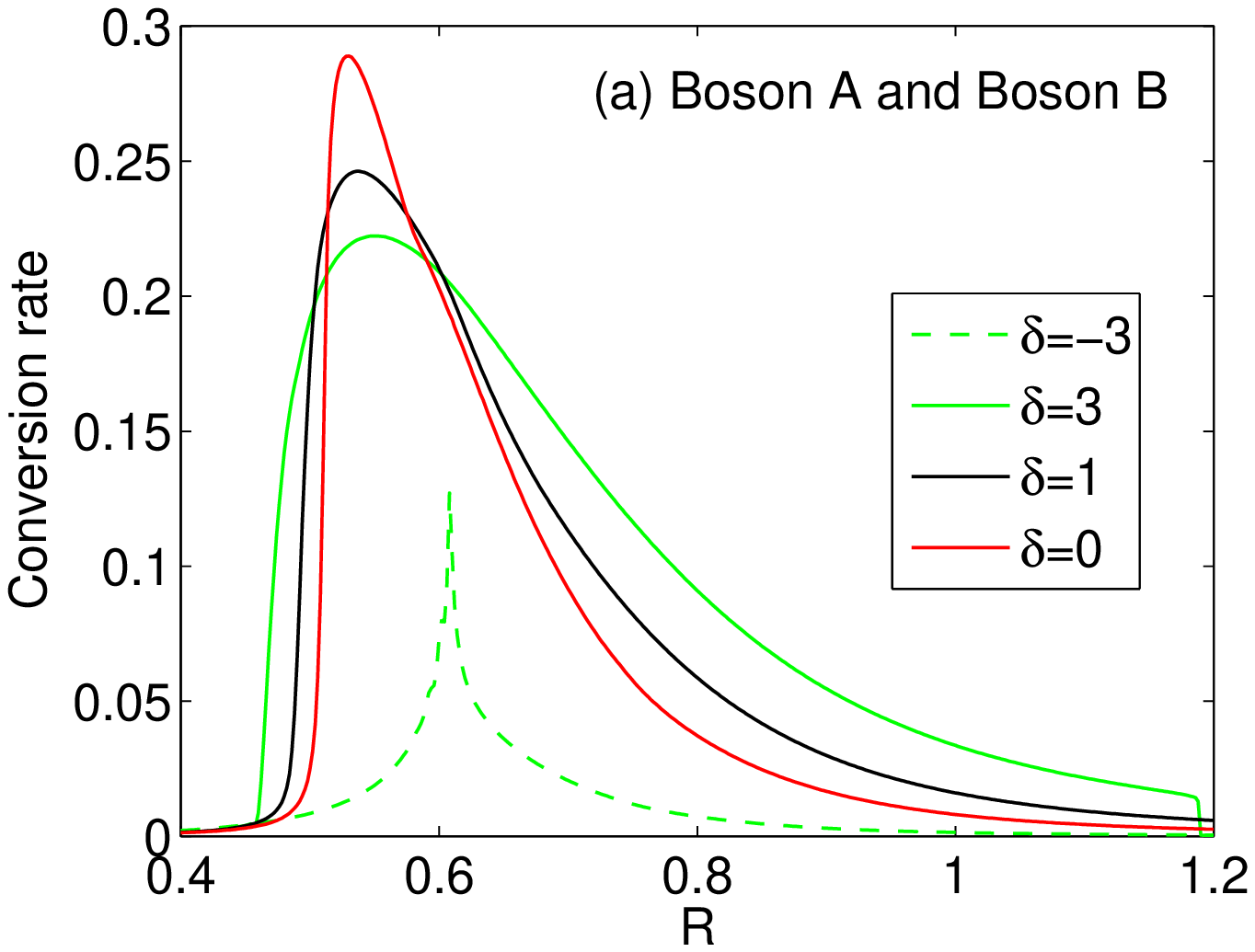}
\includegraphics[width=8.5cm]{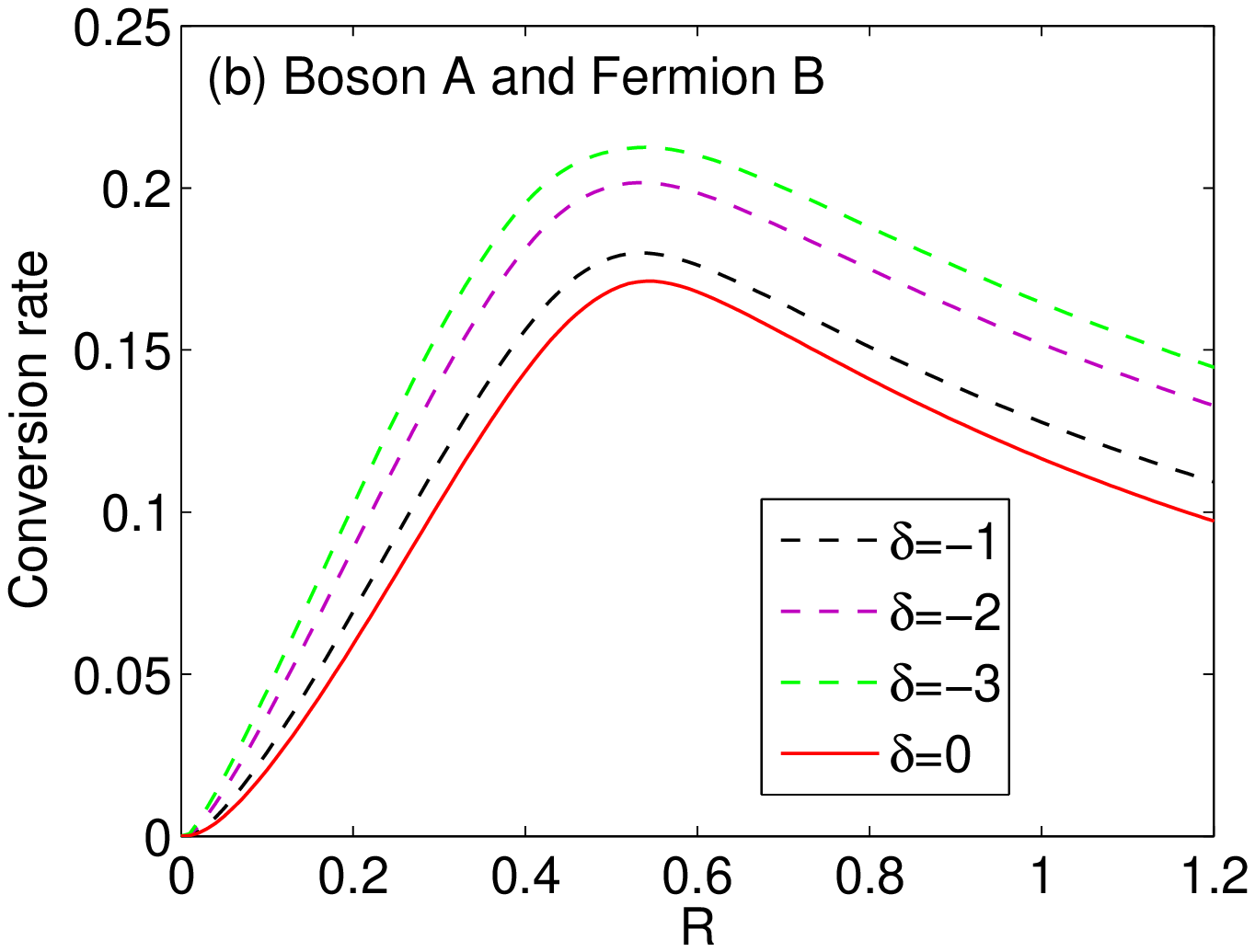}
\caption{(Color online) Final dimer population
$|\psi_{ab}(t=\infty)|^2$ as the function of R for (a) the Bose case
and (b) the Fermi case for several values of $\delta$.}
\end{figure}

Finally, we remark that we can follow a similar approach to study
the collective multi-molecular reactions $2AB \rightarrow A_2 + B_2$
and $2A_2 \rightarrow A_3 + A$. In the first case we find that the
steady-state CPT values of the product dimer $A_2$ or $B_2$ are the
same as that of $AB$ in the $A+B_2\rightarrow AB+B$ reaction for
$\cal {R}$=1/2, see Table I. For the reaction $2A_2 \rightarrow A_3
+ A$,  we obtain the equations of motion
\begin{align}
\dot{\psi}_{ a}=&2i \sum_{ j} \chi_{{ a,j}} |\psi_{ j}|^2
\psi_{{ a}}-i\Omega\psi_{ a_3}^*\psi_{ t};\nonumber\\
\dot{\psi}_{ a_2}=&2i \sum_{ j} \chi_{{ a_2,j}} |\psi_{ j}|^2
\psi_{ a_2}+2i\lambda\psi_{ a_2}^*\psi_{ t};\nonumber\\
\dot{\psi}_{ a_{3}}=&2i \sum_{ j} \chi_{ a_3,j} |\psi_{ j}|^2
\psi_{ a_3}+i(\Delta+\delta)\psi_{ a_3}
-i\Omega\psi_{ a}^*\psi_{ t},\nonumber\\
\dot{\psi}_{ t}=&2i \sum_{ j} \chi_{{ t,j}} |\psi_{ j}|^2 \psi_{
t}+(i\delta-\gamma)\psi_{ t}+i\lambda\psi_{ a_2}\psi_{ a_2}\nonumber\\
&-i\Omega\psi_{ a_3}\psi_{ a},
\end{align}
from which it is straightforward to show that the ``two-photon"
resonance condition is
\begin{eqnarray}
\Delta\!\!\!&=&\!\!\!-\delta+\!(3\chi_{ a_2,a}-2\chi_{
a_t,a})|\psi_a^s|^2\!+\!(3\chi_{ a_2,a_2}-2\chi_{
t,a_2})|\psi_{a_2}^s|^2\nonumber\\
&&+(3\chi_{ a_2,a_3}-2\chi_{ t,a_3})|\psi_{a_3}^s|^2.
\end{eqnarray}
Introducing a steady-state ansatz similar to Eqs.~(12), we
can then calculate the CPT value of the trimer population as
\begin{align}
N_{ a_{3},a}^{s}=\frac{1}{2(2+\Omega/\lambda)}.
\end{align}
We have carried out numerical simulations for this case and found
results similar to those sumarized in Fig.~3.

\section{Conclusion}

In conclusion, we have shown that in quantum-degenerate ultracold
atomic and molecular systems the abstraction reaction
$A+B_2\rightarrow AB+B$ can be fundamentally different from the
familiar atom-molecule combination. The quantum noise that
triggers this reaction leads to the creation of strongly
correlated dimer-atom pairs \cite{Wu}. In addition, a generalized
atom-molecule dark state existing in this system can significantly
enhance the creation of boson-boson or fermi-fermi dimer-atom
pairs. Our study opens up a new direction in association
resonances and thus a fascinating new regime of superchemistry. In
particular it can be generalized to the analysis of more
multi-molecular matter-wave reactions, such as the reaction
$2AB\rightarrow A_2+B_2$ or the creation of a quantum-entangled
atom-molecule laser $A+A+A \rightarrow A+A_2$ [12, 39, 62],
controlled even by purely optical means.

Future work will investigate the unique ``superchemistry" effects
of ultra-selectivity or confinement-induced stability on the
bimolecular reactions, the collective abstraction reaction in a
double-well potential or an optical lattice, and the possible
applications of these kinds of reactions in realizing
laser-catalyzed atomic spin mixing in a spinor-1 Bose condensate
\cite{PD, 09}. In addition, a complete analysis of collisional
effects \cite{3body,Ni} will be considered. While experiments
along the line of this analysis promise to be challenging, recent
progress in quantum degenerate chemistry \cite{Winkler,PD,Ling}
and in the manipulation of atom-molecule systems
\cite{trimer,tetramer,triple,08,Nielsen,Papp} indicates that
achieving this goal should become possible in the not too distant
future.

\acknowledgments One of the authors (H.J.) is grateful for
useful discussions with Yajing Jiang and Yuangang Deng. P.M is
supported by the US Office of Naval Research, by the US National
Science Foundation, and by the US Army Research Office. J.C. is
supported by the National Natural Science Foundation of China
(Grants No. 10404031 and No. 10774047) and the SCUT BaiRen
Program. H.J. is supported by the National Natural Science
Foundation of China (Grant No. 10874041), by the Program for New
Century Excellent Talents in University, and by the Henan Province
Talented Youth Program.


\begin{thebibliography}{99}

\bibitem{krems}
K. Bongs and K. Sengstock, Rep. Prog. Phys. {\bf 67}, 907 (2004); R.
Krems, Int. Rev. Phys. Chem. {\bf 24}, 99 (2005).

\bibitem{Robert}
J. L. Roberts, N. R. Claussen, S. L. Comish, E. A. Donley, E. A.
Comell, and C. E. Wieman, Phys. Rev. Lett. {\bf 86}, 4211 (2001).

\bibitem{FR 1}
M. Mackie, K. A. Suominen, and J. Javanainen, Phys. Rev. Lett. {\bf
89}, 180403 (2002).

\bibitem{Greiner}
M. Greiner, C. A. Regal, and D. S. Jin, Nature (London) {\bf 426},
537 (2003).

\bibitem{Kerman}
A. J. Kerman, J. M. Sage, S. Sainis, T. Bergeman, and D. DeMille,
Phys. Rev. Lett. {\bf 92}, 153001 (2004).

\bibitem{Danzl}
J. G. Danzl, E. Haller, M. Gustavsson, M. J. Mark, R. Hart, N.
Bouloufa, O.Dulieu, H. Ritsch, and H. C. N$\ddot{a}$gerl, Science
{\bf 321}, 1062 (2008).

\bibitem{PA 1}
J. Javanainen and M. Mackie, Phys. Rev. A. {\bf 58}, R789 (1998).

\bibitem{PA 2}
J. Javanainen and M. Mackie, Phys. Rev. A {\bf 59}, R3186 (1999).

\bibitem{PA 3}
M. Ko\u{s}trun, M. Mackie, R. C\^{o}te, and J. Javanainen, Phys.
Rev. A {\bf 62}, 063616 (2000).

\bibitem{Paul}
T. K$\ddot{o}$hler, K. G$\acute{o}$ral, and P. S. Julienne, Rev.
Mod. Phys. {\bf 78}, 1311 (2006);K. M. Jones, E. Tiesinga, P. D.
Lett, and P. S. Julienne, ${\it ibid.}$ {\bf 78}, 483 (2006).

\bibitem{Lang}
 F. Lang, K. Winkler, C. Strauss, R. Grimm, and J. Hecker Denschlag, Phys. Rev. Lett. {\bf 101}, 133005 (2008);
J. Deiglmayr, A. Grochola, M. Repp, K. M$\ddot{o}$rtlbauer, C.
Gl$\ddot{u}$ck, J. Lange, O. Dulieu, R. Wester, and M.
Weidem$\ddot{u}$ller, Phys. Rev. Lett. {\bf 101}, 133004 (2008);  M.
Viteau,  A. Chotia, M. Allegrini, N. Bouloufa, O. Dulieu,
D.Comparat, and P. Pillet, Science {\bf 321}, 232 (2008);  J. G.
Danzl, E. Haller, M. Gustavsson, M. J. Mark, R. Hart, N. Bouloufa,
O.Dulieu, H. Ritsch, and H. C. N$\ddot{a}$gerl, Science {\bf 321},
1062 (2008);  K. K. Ni, S. Ospelkaus, M. H. G. de Miranda, A. Pe'er,
B. Neyenhuis, J. J. Zirbel, S. Kotochigova, P. S. Julienne, D. S.
Jin, and J. Ye, Science {322}, 231 (2008).

\bibitem{trimer}
T. Kraemer, M. Mark, P. Waldburger, J. G. Danzl, C. Chin, B.
Engeser, A. D. Lange, K. Pilch, A. Jaakkola, H. C. N$\ddot{a}$gerl,
and R. Grimm, Nature (London) {\bf 440}, 315 (2006).

\bibitem{Efimov}
V. Efimov, Phys. Lett. {\bf 33B}, 563 (1970).

\bibitem{tetramer}
 C. Chin, T. Kraemer, M. Mark, J. Herbig, P.Waldburger, H. C. Nagerl, and R.Grimm, Phys. Rev. Lett. {\bf 94}, 123201 (2005).

\bibitem{Ling}
H.-Y. Ling, H. Pu, and B. Seaman, Phys. Rev. Lett. {\bf 93}, 250403
(2004).

\bibitem{Pu}
H. Pu, P. Maenner, W. Zhang, and H.-Y. Ling, Phys. Rev. Lett.
{\bf98}, 050406 (2007); H. Jing, F. Zheng, Y. Jiang, and Z. Geng,
Phys. Rev. A {\bf 78}, 033617 (2008).

\bibitem{Ling 1}
H.-Y. Ling, P. Maenner, W. Zhang, and H. Pu, Phys. Rev. A {\bf 75},
033615 (2007).

\bibitem{nonlinear 1}
S. J. J. M. FKokkelmans, H. M. J. Vissers, and B. J. Verhaar, Phys.
Rev. A {\bf63}, 031601 (2001).

\bibitem{nonlinear 2}
E. A. Doaley, N. R. Claussen, S. T. Thompson, and C. E. Wieman,
Nature (London), {\bf  417},529 (2002).

\bibitem{F}
M. Mackie, Phys. Rev. A {\bf66}, 043613 (2002).

\bibitem{STIRAP 1}
K. Bergmann, H. Theuer, and B. Shore, Rev. Mod. Phys. {\bf 70}, 1003
(1998).

\bibitem{STIRAP 2}
N. V. Vitanov, T. Halfmann, B. W. Shore, and K. Bergmann, Annu. Rev.
Phys. Chem. {\bf 52}, 763 (2001).

\bibitem{STIRAP 3}
M. Mackie and J. Javanainen, Phys. Rev. A {\bf 60}, 3174 (1999).

\bibitem{STIRAP 4}
M. Mackie, R. Kowalski, and J. Javanainen, Phys. Rev. Lett. {\bf
84}, 3803 (2000).

\bibitem{Winkler}
K. Winkler, G. Thalhammer, M. Theis, H. Ritsch, R. Grimm, and J. H.
Denschlag, Phys. Rev. Lett. {\bf 95}, 063202 (2005).

\bibitem{Heinzen}
D. J. Heinzen, R. Wynar, P. D. Drummond, and K. V. Kheruntsyan,
Phys. Rev. Lett. {\bf 84}, 5029 (2000); J. J. Hope and M. K. Olsen,
${\it ibid.}$ {\bf 86}, 3220 (2001); J. J. Hope, Phys. Rev. A
{\bf64}, 053608 (2001); J. J. Hope, M. K. Olsen, and L. I. Plimak,
$ibid.$ {\bf63}, 043603 (2001).

\bibitem{PD}
M. G. Moore and A. Vardi, Phys. Rev. Lett. {\bf 88}, 160402 (2002);
 A. Vardi and M. G. Moore, ${\it ibid.}$ {\bf 89}, 090403 (2002).

\bibitem{Bongs}
 C. Ospelkaus, S. Ospelkaus, L. Humbert, P. Ernst, K. Seng. Stock, and K. Bongs, Phys. Rev. Lett. {\bf 97}, 120402
(2006).

\bibitem{Vardi}
I. Tikhonenkov and A. Vardi, ${\it ibid.}$ {\bf 98}, 080403 (2007).

\bibitem{olavi}
O. Dannenberg, M. Mackie, and K-A. Suominen, Phys. Rev. Lett.
{\bf91}, 210404 (2003).

\bibitem{bodo}
E. Bodo, F. A. Gianturco, N. Balakrishnan, and A. Dalgarno, J. Phys.
B {\bf 37}, 3641 (2004).

\bibitem{che}
 L. Che, Z. F. Ren, X. G. Wang, W. R. Dong, D. X. Dai, X. Y. Wang, D. H. Zhang, X. M. Yang, L.S. Sheng, G. L. Li, H. J. Werner, F. Lique, and M. H. Alexander, Science {\bf 317}, 1061 (2007).

\bibitem{disp}
M. Shapiro and P. Brumer, {\it Principles of the Quantum Control of
Molecular Processes} (Wiley, New York, 2003); Phys. Rev. Lett. {\bf
77}, 2574 (1996); C. A. Arango, M. Shapiro, and P. Brumer, ${\it
ibid.}$ {\bf 97}, 193202 (2006).

\bibitem{QO}
P. Meystre and M. Sargent  IIII, {\it Elements of Quantum Optics}
(4th Edition, Springer-Verlag, 2007).

\bibitem{Uys}
M. G. Moore and P. Meystre, Phys. Rev. Lett. {\bf 83}, 5202
(1999); H. Pu and P. Meystre, ${\it ibid.}$ {\bf 85}, 3987 (2000);
H. Uys and P. Meystre, Phys. Rev. A {\bf 75}, 033805 (2007).

\bibitem{amplifier}
C. P. Search and P. Meystre, Phys. Rev. Lett. {\bf 93}, 140405
(2004).

\bibitem{Jing}
H. Jing, J. Cheng, and P. Meystre, Phys. Rev. Lett. {\bf 99},
133002 (2007); ${\it ibid.}$ {\bf 101}, 073603 (2008).

\bibitem{3body}
B.Borea, J. W. Dunn, V. Kokoouline, and C. H. Greene, Phys. Rev.
Lett. {\bf91}, 070404 (2003).

\bibitem{footnote1} We note that the trimer formation via an
atom-dimer resonance is actively pursued in experiments experiments.
The nonlinear photodissociation of trimers, typically induced by a
narrow frequency continuous-wave laser, and the manipulation of the
dissociation channels is also studied extensively (see, e.g., V.
Zhaunerchyk ${\it et~al.}$, Phys. Rev. Lett. {\bf 98}, 223201 (2007)
or S. Jung ${\it et~ al.}$, J. Phys. B {\bf 39}, S1085 (2006)). The
Franck-Condon factor for this transition can be calculated, at least
in principle, by a diatomics-in-molecules description or by other
simulations of potential energy surfaces (see, e.g., B. L.
Grigorenko, A. V. Nemukhin, and V. A. Apkarian, Chem. Phys. {\bf
219}, 161 (1997)).

\bibitem{rogue1}
J. Javanainen and M. Mackie, Phys. Rev. Lett. {\bf 88}, 090403
(2002).

\bibitem{Wu}
Y. Shi and Q. Niu, Phys. Rev. Lett. {\bf 96}, 140401 (2006).

\bibitem{twin} K. V. Kheruntsyan, Phys. Rev. Lett. {\bf
96}, 110401 (2006); W. Zhang, C. P. Search. H. Pu, P. Meystre, and
E. M. Wright, ${\it ibid.}$ {\bf 90}, 140401 (2003).

\bibitem{footnote2} Dimer-atom entanglement can also be studied in the general
case by applying for example the positive-$P$ technique in quantum
optics \cite{Heinzen} or other methods \cite{Wu}.

\bibitem{STIRAP 5}
F. T. Hioe and J. H. Eberly, Phys. Rev. A {\bf 29}, 1164 (1984).

\bibitem{STIRAP 6}
J. Oreg, F. T. Hioe, and J. H. Eberly, Phys. Rev. A {\bf 29}, 690
(1984).

\bibitem{STIRAP 7}
J. Klein, F. Beil, and T. Halfmann,  Phys. Rev. Lett. {\bf 99},
113003 (2007).

\bibitem{M.N}
M. N. Kobrak and S. A. Rice, Phys. Rev. A {\bf 57}, 2885 (1998).

\bibitem{Z.Kis}
Z. Kis, and S. Stenholm, Phys. Rev. A {\bf 64}, 063406 (2001).

\bibitem{adiabatic condition}
K. P. Marzlin and B. C. Sanders, Phys. Rev. Lett. {\bf 93}, 160408
(2004).

\bibitem{lens}
G. Modugno, M. Modugno, F. Riboli, G. Roati, and M. Inguscio , Phys.
Rev. Lett. {\bf 89}, 190404 (2002).


\bibitem{scattere length}
M. Mackie, K. H$\ddot{a}$rk$\ddot{o}$nen, A. Collin, K. A. Suominen,
and J. Javanainen, Phys. Rev. A {\bf 70}, 013614 (2004).


\bibitem{rogue2}
O. Dannenberg and M. Mackie, Phys. Rev. A {\bf 74}, 053601 (2006).

\bibitem{rogue3}
M. Macki, M. Fenty, D. Savage, and J. Kesselman, Phys. Rev. Lett.
{\bf 101}, 040401 (2008).

\bibitem{later} The details of the calculations will be given
elsewhere.

\bibitem{quench}
M. T. Cvita$\breve{s}$, P. Sold$\acute{a}$n, J. M. Hutson, P.
Honvault, and J. M. Launay, Phys. Rev. Lett. {\bf 94}, 200402
(2005); M. T. Cvita$\breve{s}$, P. Sold$\acute{a}$n, J. M. Hutson,
P. Honvault, and J. M. Launay, ${\it ibid.}$ {\bf 94}, 033201
(2005).

\bibitem{LuLi}
L.-H. Lu and Y.-Q. Li, Phys. Rev. A {\bf 76}, 053608 (2007).

\bibitem{Parkin}
A. S. Parkins and D. F. Walls, Phys. Rep. {\bf303}, 1 (1998).

\bibitem{Salerno}
M. Salerno, Phys. Rev. A, {\bf72}, 063602 (2005).

\bibitem{meystre}
M. G. Moore and P. Meystre, Phys. Rev. Lett. {\bf 86}, 4199
(2001).

\bibitem{08}
X. Li, G. A. Parker, P. Brumer, I. Thanopulos, and W. Shapiro, Phys.
Rev. Lett. {\bf 101}, 043003 (2008)).

\bibitem{zwier}
M. W. Zwierlein, A. Schirotzek, H. Schunck, and W. Ketterle, Science
{\bf 311}, 492 (2006).

\bibitem{amlaser} B. Borca, J. W. Dunn, V. Kokoouline, and C. H.
Greene, Phys. Rev. Lett. {\bf 91}, 070404 (2003).


\bibitem{09} H. Jing , Y. Jiang, W. Zhang, and P. Meystre,New J. Phys.{\bf 10},123005(2008).

\bibitem{Ni} J. P. D'Incao and B. D. Esry, Phys. Rev. Lett. {\bf 94}, 213201
(2005); E. Braaten and H. W. Hammer, Phys. Rep. {\bf 428}, 259
(2006).

\bibitem{triple}
M. Taglieber, A. C. Voigt, T. W. Hansch, and K. Dieckmann, Phys.
Rev. Lett. {\bf 100}, 010401 (2008).

\bibitem{Nielsen}
E. Nielsen, H. Suno, and B. D. Esry, Phys. Rev. A {\bf 66}, 012705
(2002); M. Mackie, O. Dannenberg, J. Piilo, K. A. Suominen, and J.
Javanainen, Eur. Phys. J. D {\bf 31}, 273 (2004).

\bibitem{Papp}
 P. Staanum, S. D. Kraft, J. Lange, R. Wester, and M. Weidem$\ddot{u}$ller, Phys. Rev. Lett. {\bf 96}, 023201
(2006); S. B. Papp and C. E. Wieman, ${\it ibid.}$ {\bf 97}, 180404
(2006).








\end{thebibliography}
\end{document}